\documentclass[
    aps,
    prd,preprintnumbers,
    reprint,
    superscriptaddress,
    nofootinbib,
    longbibliography
]{revtex4-2}

\usepackage{amsmath,amssymb}
\usepackage{bm}
\usepackage{graphicx}
\usepackage[dvipsnames]{xcolor}
\usepackage{hyperref}
\hypersetup{hidelinks}
\newcommand{\be}{\begin{equation}}
\newcommand{\ee}{\end{equation}}
\newcommand{\ba}{\begin{eqnarray}}
\newcommand{\ea}{\end{eqnarray}}

\newcommand{\circled}[1]{\textcircled{\textbf{#1}}}

\begin{document}

\title{Topological Multi-Natural Inflation}
\author{Siao Yang Chai}
\affiliation{
Department of Physics, Tohoku University, Sendai, Miyagi 980-8578, Japan
}
\author{Fuminobu Takahashi}
\affiliation{
Department of Physics, Tohoku University, Sendai, Miyagi 980-8578, Japan
}

\preprint{TU-1330}
\date{\today}

\begin{abstract}
We investigate topological inflation in minimal multi-natural inflation, in which the axion potential consists of two cosine terms. Focusing on the high-scale regime, we perform a systematic parameter scan for representative ratios of the two periods and identify regions consistent with current constraints on the scalar spectral index and tensor-to-scalar ratio, including those from the Atacama Cosmology Telescope. The interplay of the two harmonics allows departures from the quartic hilltop limit and accommodates the higher spectral index favored by recent cosmological data combinations. We characterize the hilltop and near-inflection-point branches and analyze the associated vacuum structure and domain-wall properties. We also discuss a symmetry-breaking origin of the required domain walls when the inflaton is the angular component of a complex scalar field. We identify regions in which the observational constraints and the adopted criteria for topological inflation are simultaneously satisfied, including regions with a sub-Planckian axion decay constant.
Our results show that the observational tension encountered in the minimal quartic double-well model need not extend to topological inflation in a minimal multi-natural axion potential.
\end{abstract}

\maketitle



\section{Introduction}
\label{sec:introduction}

Inflation provides a solution to the horizon and flatness problems of
hot Big Bang cosmology~\cite{Guth:1980zm,Starobinsky:1980te,Sato:1981qmu}.
Its own initial conditions, however, require an explanation.
In the original picture of chaotic inflation~\cite{Linde:1983gd},
one considers Planck-size regions with kinetic, gradient, and
potential energy densities of order the Planck density.
Regions in which the potential energy happens to dominate can
begin to inflate, provided the potential is sufficiently flat.

Topological inflation instead relies on defects that keep the
inflaton near a potential barrier~\cite{Vilenkin:1994pv,Linde:1994hy}.
In a domain wall, the field necessarily crosses the barrier,
and the potential-dominated core can undergo
inflation~\cite{Sakai:1995nh}.
A standard domain-wall realization involves physically distinct degenerate vacua, a mechanism to populate them in different spatial regions, and a potential-dominated wall core with a width comparable to or larger than the local Hubble radius. A simple example employs a $Z_2$-symmetric double-well potential, with vacuum degeneracy enforced by $\phi\to-\phi$~\cite{Izawa:1998rh,Kawasaki:2023mjm}. For commonly studied double-well potentials, the wall-width condition typically requires an inflaton vacuum expectation value of order the Planck scale or larger~\cite{Sakai:1995nh,Kawasaki:2000tv,Sakai:2003st,Choi:2022ssv}.\footnote{
Topological inflation can also be driven by the Standard Model Higgs field, explaining the origin of the criticality~\cite{Hamada:2014raa}.}

In the minimal quartic double-well model, the potential decreases quadratically away from its maximum, making it difficult to obtain a scalar spectral index $n_s$ sufficiently close to unity while keeping the tensor-to-scalar ratio $r$ small~\cite{Kawasaki:2023mjm}.  Joint analyses of Atacama Cosmology Telescope (ACT) Data Release 6, Planck, cosmic microwave background (CMB) lensing, and baryon acoustic oscillation data from the Dark Energy Spectroscopic Instrument favor a larger $n_s$ than the Planck 2018 baseline, sharpening the tension for this model~\cite{Planck:2018vyg,AtacamaCosmologyTelescope:2025blo,AtacamaCosmologyTelescope:2025nti}.

An axion provides a well-motivated inflaton candidate whose approximate shift symmetry protects the flatness of its potential~\cite{Freese:1990rb}. Natural inflation contains a single cosine term, whereas multi-natural inflation allows several periodic contributions~\cite{Czerny:2014wza,Czerny:2014xja}. Already the minimal two-cosine potential admits a variety of shapes through the relative amplitude, period, and phase of the two terms, including hilltops and near-inflection-point regions. This flexibility motivates exploring whether the inflationary observables and the requirements for topological inflation can be satisfied simultaneously.

The axion interpretation also suggests a mechanism for
domain-wall formation. For example, the inflaton can arise
as the angular component of a complex scalar field whose
global $U(1)$ symmetry is spontaneously broken in an earlier
phase transition. During this transition, the angular field
takes independent values in causally disconnected
regions~\cite{Kibble:1976sj}.
As the axion potential subsequently becomes important,
the field settles into its minima, forming domain walls
between regions that reach distinct degenerate vacua.

In this paper, we investigate topological inflation in minimal multi-natural inflation, with particular attention to its compatibility with current CMB observations.
We consider a potential consisting of two cosine terms
and explore its parameter space for representative choices
of their period ratio.
Varying the relative amplitude and phase of the two terms,
as well as the axion decay constant, we identify hilltop and
near-inflection-point regimes consistent with current CMB
constraints on $n_s$ and $r$.
We then examine the vacuum structure and domain-wall
properties of these potentials and discuss a possible
symmetry-breaking origin of the walls.
We identify parameter regions that satisfy both the observational constraints and the adopted criteria for topological inflation. These include configurations with sub-Planckian axion decay constants, as well as reflection-symmetric potentials with nearly flat shoulders.

The rest of this paper is organized as follows.
In Sec.~\ref{sec:model}, we introduce the two-cosine
axion potential, its characteristic features, and the
inflationary dynamics.
Section~\ref{sec:scan} presents the parameter scan
and identifies regions consistent with CMB constraints,
including hilltop and near-inflection-point regimes.
In Sec.~\ref{sec:topological}, we discuss the vacuum
structure and domain-wall formation, and examine
the conditions for topological inflation in the
observationally viable parameter space.
We conclude in Sec.~\ref{sec:conclusions}.


\section{Minimal multi-natural inflation}
\label{sec:model}

\subsection{Potential and axion realization}
\label{sec:potential}

In natural inflation, the inflaton is an axion whose approximate
shift symmetry is explicitly broken by a periodic
potential~\cite{Freese:1990rb},
\begin{equation}
\label{natural}
V(\phi)=\Lambda^4\left[1-\cos\left(\frac{\phi}{f}\right)\right],
\end{equation}
where $\Lambda$ sets the potential scale and $f$ is the axion
decay constant. The single-cosine potential is now strongly
disfavored by CMB observations, as its predictions for the
scalar spectral index $n_s$ and tensor-to-scalar ratio $r$
cannot simultaneously satisfy the observational constraints
for conventional inflationary histories~\cite{Stein:2021uge}.

Multi-natural inflation can alleviate this tension by introducing multiple sources of explicit shift-symmetry breaking, which generate cosine terms with different periodicities~\cite{Czerny:2014wza,Czerny:2014xja}. The interplay of these terms provides additional freedom to adjust the slope and curvature of the inflaton potential.
We consider the minimal two-cosine potential,
\begin{align}
\label{multinatural}
V(\phi)={}&C-\Lambda^4\cos\left(\frac{\phi}{f}\right)
-B\Lambda^4\cos\left(\frac{\phi}{Af}+\theta\right).
\end{align}
Here $A>0$ and $B>0$ are the ratios of the periods and
amplitudes of the second term to the first, respectively,
$\theta$ is their relative phase, and $C$ is chosen so that
the global minimum has zero potential energy.
The limit $B=0$ recovers natural inflation.

A simple origin for the two terms is obtained by identifying
$\phi$ with the angular component of a complex scalar~\cite{Czerny:2014wza,Czerny:2014xja},
\begin{equation}
\label{eq:axion-embedding}
\Phi=\frac{F+\rho}{\sqrt{2}}\,e^{i\phi/F}.
\end{equation}
Here $F$ is the symmetry-breaking scale, and the radial mode
$\rho$ is assumed to be heavy and stabilized at zero during
inflation. Thus, $\phi$ is a Nambu-Goldstone boson associated with the spontaneous breaking of a global U(1) symmetry under which $\Phi$ has a unit charge.
We introduce $n_q$ vectorlike pairs of Weyl fermions
$(q_i,\bar q_i)$ transforming as the fundamental and
antifundamental representations of a hidden $SU(N_1)$,
and $n_Q$ analogous pairs $(Q_j,\bar Q_j)$ under a
distinct hidden $SU(N_2)$.
Their Yukawa interactions are
\begin{align}
\label{eq:yukawa}
\mathcal{L}_Y={}&-\sum_{i=1}^{n_q}y_i\Phi q_i\bar q_i
-\sum_{j=1}^{n_Q}Y_j\Phi Q_j\bar Q_j+\mathrm{h.c.},
\end{align}
where $y_i$ and $Y_j$ are Yukawa couplings.
The fermions are vectorlike under the hidden gauge groups
but chiral under the global $U(1)$ symmetry, with charges
assigned so that the Yukawa interactions are invariant.
This $U(1)$ is explicitly broken by anomalies with the
two hidden gauge groups, with anomaly coefficients
$n_q$ and $n_Q$, respectively.
After integrating out the heavy fermions, nonperturbative
effects in the two gauge sectors generate the two cosine
terms in Eq.~\eqref{multinatural}, with
\begin{equation}
\label{eq:anomaly-matching}
f=\frac{F}{n_q},\qquad
A=\frac{n_q}{n_Q},\qquad
B=\frac{\Lambda_2^4}{\Lambda_1^4},
\end{equation}
where $\Lambda=\Lambda_1$ and $\Lambda_{1,2}^4$ are the
nonperturbative potential amplitudes of the two sectors.
In this realization, $A$ is rational and the two terms share
a common period.

With both cosine terms present, the integer anomaly coefficients
imply a residual discrete shift symmetry,
\begin{equation}
\label{eq:discrete-shift}
\phi\longrightarrow\phi+\frac{2\pi F}{d},\qquad
d\equiv\gcd(n_q,n_Q).
\end{equation}
The physical field identification is
$\phi\sim\phi+2\pi F$.\footnote{
The periodicity of the axion potential alone does not determine
the physical identification of field values.
Here $\phi\sim\phi+2\pi F$ follows
from the interpretation of $\phi/F$ as the phase of $\Phi$.
}
Thus, the potential generically has $d$ physically distinct,
exactly degenerate global minima within one physical period.
For $d=1$, the discrete shift reduces to the physical field
identification, and the global vacuum is generically unique. 
Still, an axion domain wall can exist as a configuration
in which $\phi$ winds by $2\pi F$ around a cosmic string, connecting the same physical vacuum on both sides.
In this case, each string is attached to
a single domain wall, and the resulting string--wall
system is unstable.
Nevertheless, once the axion potential becomes dynamically
relevant and the wall forms, topological inflation can
begin in a sufficiently broad wall core, even while
the wall remains attached to a string.
Additional degenerate minima can arise from reflection
symmetry, as discussed below.

\subsection{Potential structure relevant to topological inflation}
\label{sec:potential-structure}

The inflationary observables and the conditions for
topological inflation probe different aspects of the
axion potential.
The scalar spectral index and tensor-to-scalar ratio
are primarily determined by the potential along the
slow-roll trajectory, typically between a local maximum
or a nearly flat shoulder and a minimum.
In contrast, the formation of domain walls depends on
the vacuum structure over the full field range.
Whether a domain wall can undergo inflation further
depends on the potential shape near its core.
For reflection-symmetric wall configurations, the
curvature at the potential maximum is particularly
important, as shown by the analysis of self-gravitating
domain walls by Ref.~\cite{Bonjour:1999kz}.
We discuss the corresponding conditions for topological
inflation in Sec.~\ref{sec:wall-conditions}.

To describe the potential shape,
define $X\equiv\phi/f$ and write
\begin{equation}
\label{eq:dimensionless-potential}
U(X)\equiv\frac{V(\phi)}{\Lambda^4}
=c-\cos X-B\cos\left(\frac{X}{A}+\theta\right),
\end{equation}
where $c=C/\Lambda^4$. The derivatives are
\begin{align}
U_X&=\sin X+\frac{B}{A}\sin\left(\frac{X}{A}+\theta\right),
\notag\\
U_{XX}&=\cos X+\frac{B}{A^2}\cos\left(\frac{X}{A}+\theta\right).
\label{eq:potential-derivatives}
\end{align}
The relative contributions of the second harmonic to the
slope and curvature scale differently, as $B/A$ and $B/A^2$.
Combining the conditions $U_X=U_{XX}=0$ gives
\begin{equation}
\label{eq:stationary-degeneracy}
B^2=A^2\sin^2X+A^4\cos^2X.
\end{equation}
Hence, for $0<A<1$, such simultaneous cancellations require
$A^2\leq B\leq A$.

\subsubsection{Hilltop regime}
\label{sec:hilltop-regime}

In the hilltop regime, inflation takes place near a local
maximum. A particularly flat configuration is obtained by
aligning a maximum of the first contribution with a minimum
of the second. Let $X_h$ and $\theta_h$ satisfy
\begin{equation}
\label{eq:hilltop-alignment}
X_h=(2k+1)\pi,\qquad
\frac{X_h}{A}+\theta_h=2\ell\pi,
\end{equation}
where $k,\ell\in\mathbb{Z}$.
At $\theta=\theta_h$, the potential is reflection symmetric
about $X_h$, and its expansion is
\begin{align}
U(X_h+\Delta X)={}&U(X_h)
+\frac{1}{2}\left(\frac{B}{A^2}-1\right)(\Delta X)^2
\notag\\
&{}+\frac{1}{24}\left(1-\frac{B}{A^4}\right)(\Delta X)^4
+\mathcal{O}\bigl((\Delta X)^6\bigr).
\label{eq:hilltop-expansion}
\end{align}
At $B=A^2$, the quadratic term vanishes. For $0<A<1$,
the quartic coefficient is negative, realizing quartic
hilltop inflation~\cite{Czerny:2014wza,Daido:2017wwb}.
In the small-field limit, where observable inflation takes
place sufficiently close to the quartic hilltop, the leading
prediction is $n_s\simeq 1-3/N_*$,
where $N_*$ is the number of e-folds from pivot-scale horizon
exit to the end of inflation.
Departures from $B=A^2$ modify the quadratic term, while a
small phase displacement $\theta=\theta_h+\delta\theta$
generates an effective linear term in the expansion around
$X_h$, with $U_X(X_h)=(B/A)\sin\delta\theta$.
The latter is particularly effective at raising $n_s$
above the quartic hilltop prediction, since even a small
linear term can shift the field value at
horizon exit~\cite{Takahashi:2013cxa}.

\subsubsection{Near-inflection-point regime}
\label{sec:inflection-regime}

Besides the hilltop regime, the two-cosine potential can also
support slow-roll inflation along a nearly flat shoulder.
Such a configuration arises near a stationary inflection point
$X_s$, where $U_X(X_s)=U_{XX}(X_s)=0$.
At this point, the third derivative is given by
\begin{equation}
\label{eq:inflection-third-derivative}
U_{XXX}(X_s)=\left(\frac{1}{A^2}-1\right)\sin X_s.
\end{equation}
When this is nonzero, the local expansion begins at cubic order,
\begin{equation}
\label{eq:inflection-expansion}
U(X_s+\Delta X)=U(X_s)
+\frac{U_{XXX}(X_s)}{6}(\Delta X)^3
+\mathcal{O}\bigl((\Delta X)^4\bigr).
\end{equation}
Small changes in the potential parameters can produce a
nearly flat shoulder with nonzero slope, allowing the field
to roll through it while realizing a large number of
e-folds. As the parameters cross the critical configuration,
an additional local minimum and maximum can appear along
the trajectory, and the minimum can trap the classically
rolling field.\footnote{Once the inflaton is trapped in a local minimum, it could take a very large number of e-folds to get out of it due to quantum fluctuations~\cite{Kitajima:2019ibn}.}

An important feature of this regime is that the field region
responsible for observable inflation can differ from that
relevant for topological inflation.
For a reflection-symmetric potential with two distinct
degenerate minima, the domain wall is centered on the
intervening barrier, while observable slow-roll inflation
can take place near one of its nearly flat shoulders.
The scalar spectral index is then controlled by the
potential around the shoulder rather than the potential maximum.
This opens the possibility of realizing topological inflation
in the near-inflection-point regime, with observational
predictions distinct from those of quartic hilltop inflation.

\subsubsection{Field scale and wall thickness}

Topological inflation requires a sufficiently broad,
potential-dominated domain-wall core~\cite{Vilenkin:1994pv,Linde:1994hy}.
For a fixed shape of the dimensionless potential $U(X)$,
the thickness of a static domain wall scales as
$\delta_{\rm wall}\propto f/\Lambda^2$, while the Hubble parameter
associated with the potential energy near the maximum scales as
$H_{\rm inf}\propto\Lambda^2/M_{\rm Pl}$,
where $M_{\rm Pl}=(8\pi G)^{-1/2}\simeq
2.4\times10^{18}\,\mathrm{GeV}$ is the reduced Planck mass.
Consequently,
\begin{equation}
H_{\rm inf}\delta_{\rm wall}\propto\frac{f}{M_{\rm Pl}},
\end{equation}
with a proportionality factor determined by the potential
shape. Thus, the overall normalization $\Lambda$ cancels
from the wall-thickness criterion. We examine this criterion,
together with a complementary condition involving the
curvature at the barrier maximum, in
Sec.~\ref{sec:wall-conditions}.

\subsection{Inflationary dynamics and CMB observables}
\label{sec:inflation-observables}

The homogeneous inflaton dynamics can be described using
$N\equiv\ln a$ as the evolution variable.
For a canonically normalized inflaton, the Hubble
slow-roll parameter is
\begin{equation}
\label{epsH}
\epsilon_H\equiv-\frac{\dot H}{H^2}
=\frac{\phi'^2}{2M_{\rm Pl}^2},
\end{equation}
where $H=\dot a/a$, and a dot and a prime denote a derivative with
respect to cosmic time and $N$, respectively.
The inflaton equation of motion is
\begin{equation}
\label{eq:background-evolution}
\phi''+(3-\epsilon_H)\phi'
+(3-\epsilon_H)M_{\rm Pl}^2\frac{V_\phi}{V}=0,
\end{equation}
where $V_\phi=dV/d\phi$.
Inflation ends when $\epsilon_H=1$.

In the slow-roll approximation, the potential slow-roll
parameters are given by
\begin{align}
\epsilon_V&=\frac{M_{\rm Pl}^2}{2f^2}
\left(\frac{U_X}{U}\right)^2,
\\
\eta_V&=\frac{M_{\rm Pl}^2}{f^2}
\frac{U_{XX}}{U}.
\end{align}
The scalar spectral index $n_s$ and the tensor-to-scalar
ratio $r$ are then\footnote{In our numerical calculations, we use the formulae for $n_s$ and $r$ up to second order in the slow-roll parameters \cite{Kolb:1994ur}.}
\begin{align}
n_s&\simeq1-6\epsilon_{V*}+2\eta_{V*},
\\
r&\simeq16\epsilon_{V*},
\label{eq:slow-roll-observables}
\end{align}
where a star denotes evaluation at horizon exit of the
pivot scale $k_*=0.05\,\mathrm{Mpc}^{-1}$.
Thus, the ratio $f/M_{\rm Pl}$ controls both the inflationary
observables and the domain-wall width relevant for
topological inflation.

The overall scale $\Lambda$ is fixed by the CMB
normalization of the primordial scalar power spectrum.
The scalar amplitude at the pivot scale
 is
$A_s\equiv\mathcal{P}_{\mathcal{R}}(k_*)
\simeq 2.1\times10^{-9}$,
where $\mathcal{P}_{\mathcal{R}}(k)$ denotes the
dimensionless power spectrum of the comoving curvature
perturbation.
At leading order in slow roll,
\begin{equation}
A_s \simeq
\frac{\Lambda^4 U(X_*)}
     {24\pi^2 M_{\rm Pl}^4\epsilon_{V*}} .
\end{equation}

Recent CMB observations provide tight constraints on
the inflationary observables.
In the baseline $\Lambda$CDM model, the joint
Planck--ACT dataset (P-ACT), combined with CMB lensing
and DESI DR1 baryon acoustic oscillation measurements,
gives~\cite{AtacamaCosmologyTelescope:2025blo}
\begin{equation}
n_s=0.974\pm0.003
\qquad (68\%~{\rm C.L.}).
\label{eq:ns-act}
\end{equation}

The combined analysis of Planck, ACT DR6, CMB lensing, BAO, and BICEP/Keck data yields an upper bound on the tensor-to-scalar ratio of~\cite{AtacamaCosmologyTelescope:2025nti}.
\begin{equation}
r_{0.05}<0.038
\qquad (95\%~{\rm C.L.}).
\label{eq:r-bk}
\end{equation}
These observational constraints and the model
predictions in the $(n_s,r)$ plane are presented in
Fig.~\ref{ACT_ns_r_marked_Aeq0pt5}.


\section{Inflationary parameter space}
\label{sec:scan}
Before examining the conditions required for topological inflation, we first identify the parameter space in which our potential is consistent with the ACT constraints. In this section, we present a parameter scan to determine where the predicted $(n_s,r)$ values fall within the ACT $68\%$ confidence region, assuming that the pivot scale exits the horizon at $N_*=60$ (unless stated otherwise)  before the end of inflation. We focus on the period ratios $A=1/2$ and $A=1/4$, both of which preserve the fundamental periodicity of $2\pi f$. The former represents the simplest nontrivial higher-harmonic extension, while the latter allows for a richer potential structure. Comparison between these two cases provides insights into how flat regions and multiple minima emerge, which are relevant to the possibility of topological inflation. Together with the examination of the potential structure, the parameter scan allows us to identify viable parameter regions for realizing topological inflation consistent with the ACT constraints.

For each potential, we restrict the initial inflaton position to the interval between the global minimum and the nearest local maximum to its right. Starting from the midpoint of the interval, we use a bisection procedure to search for an initial position whose trajectory produces at least $70$ continuous e-folds before the end of inflation. The initial field velocity is set to $\phi_{\rm ini}'=0$. The integration is terminated at the first upward crossing of $\epsilon_H=1$, and the bisection is continued until a trajectory producing at least $70$ e-folds is found, subject to a maximum of $50$ iterations.

To reduce the computation cost for potentials whose hilltop does not itself satisfy the slow-roll condition $|\eta_V(\phi_{\rm top})|>1$, we conduct a prescreen to search for candidate regions satisfying $\epsilon_V<1,|\eta_V|<1$. If a candidate is found, we proceed with the aforementioned integration. Otherwise, as a fallback, we also test a trajectory initialized at a position one characteristic fluctuation scale away from the hilltop, 
\be
\phi_{\rm ini} = \phi_{\rm top}-\frac{H_{\rm top}}{2\pi},
\ee
for the branch rolling toward the vacuum. if neither procedure yields $70$ e-folds, the parameter point is classified as having no viable trajectory within our selected branch and initial-condition prescription.

Since $V(\phi;\theta)=V(-\phi;-\theta)$, scanning $\theta\in(-\pi,\pi)$ also accounts for the reflected trajectories approaching the corresponding minimum from the opposite direction.
The reflection-symmetric case $\theta=\pi$ is considered separately in Sec.~\ref{sec:topological} in connection with topological inflation.

\subsection{Period ratio $A=1/2$}

\begin{figure*}[ht!]
    \centering
    \includegraphics[width=\linewidth]{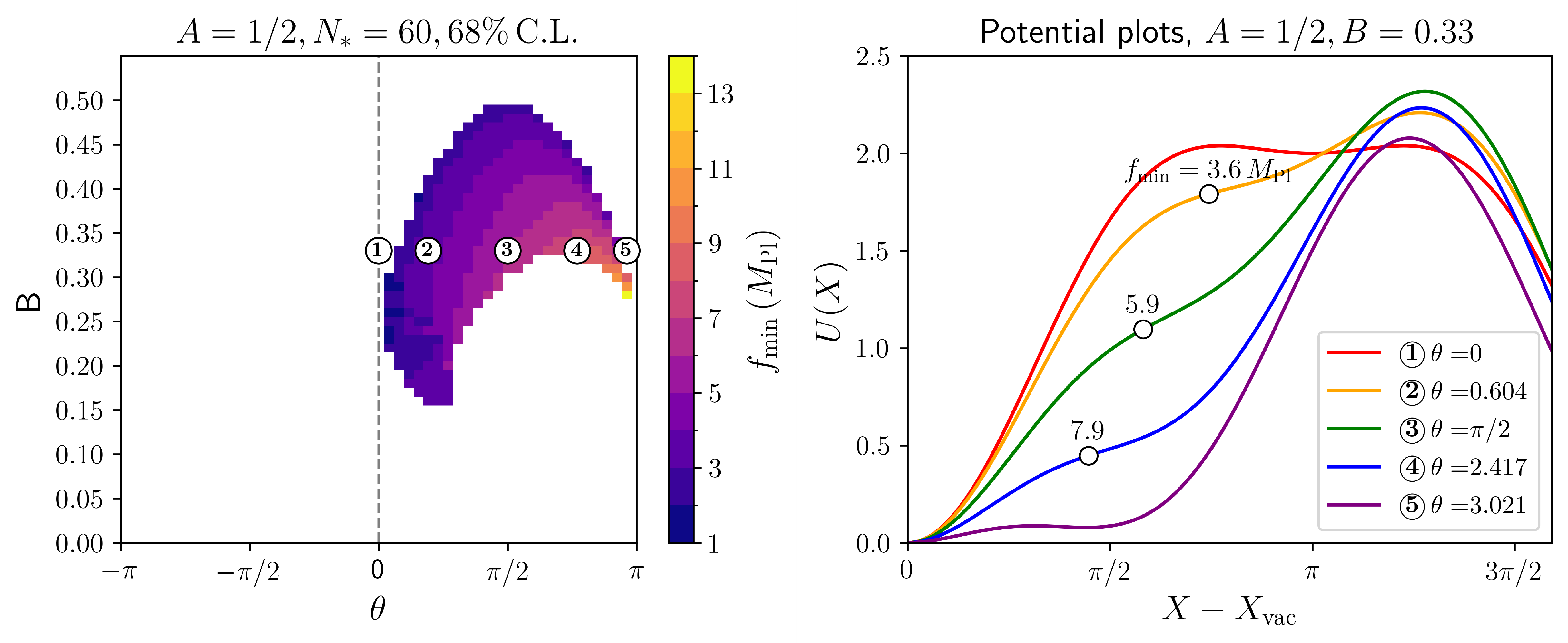}
    \caption{
        (a) Heatmap of minimum $f$ for different $(\theta,B)$ pairs at $A=1/2$, for which the predicted $(n_s,r)$ at $N_*=60$ first enter the ACT $68\%$ C.L. region. Selected points with varying $\theta$ at $B=0.33$ are marked and numbered. (b) Dimensionless inflation potentials $U(X)\equiv V(\phi)/\Lambda^4$ plotted against dimensionless field $X$, shifted by $X_{\rm vac}\equiv\phi_{\rm vac}/f$, where $\phi_{\rm vac}$ is their respective field values of vacua. Each potential corresponds to a numbered point in (a), with the $N_*=60$ horizon-exit position marked and annotated for $f=f_{\rm min}$. Horizon crossings for $\circled{1}$ and $\circled{5}$ are absent because the corresponding potentials do not produce observables within the ACT $68\%$ C.L. region for $f\leq20\,M_{\rm Pl}$.
    }
    \label{fmin_potential_combined_Aeq0pt5}
\end{figure*}

\begin{figure}[t!]
    \centering
    \includegraphics[width=\linewidth]{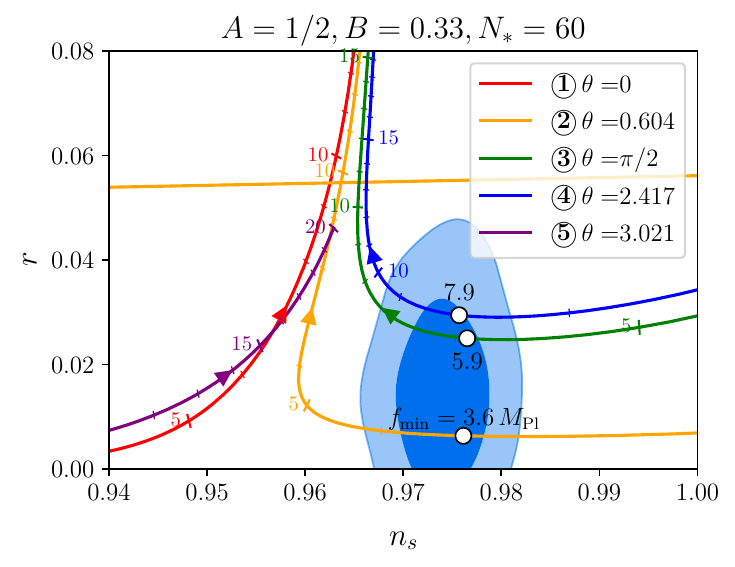}
    \caption{
    Predictions in the $(n_s,r)$ plane at $N_*=60$ for
    $(\theta,B)$ corresponding to the numbered points in Fig.~\ref{fmin_potential_combined_Aeq0pt5}(a) with fixed $A=1/2$ and $B=0.33$.
    The arrows on each trajectory point in the direction of increasing $f$ for $f\leq20 \, M_{\rm Pl}$. Minor and major ticks on each $f$-trajectory are spaced by $1$ and $5\,M_{\rm Pl}$ respectively, with the major ticks labeled by the corresponding $f$ values. The markers indicate the points along each trajectory corresponding to $f_{\rm min}$, where the predicted $(n_s,r)$ first enters the ACT 68\% C.L. region.
    }

    \label{ACT_ns_r_marked_Aeq0pt5}
\end{figure}

Fixing $A=1/2$, we calculated $(n_s,r)$ for various combinations of $(\theta,B)$, with $f$ ranging from $f=0.1\,M_{\rm Pl}$ to $20\,M_{\rm Pl}$. Fig.~\ref{fmin_potential_combined_Aeq0pt5}(a) shows the minimum values of $f$, denoted by $f_{\rm min}$, for which the predicted $(n_s,r)$ at the number of e-folds $N_* = 60$ fall within the ACT  68\% confidence-level (C.L.) region. The white region represents potentials that either cannot predict $(n_s,r)$ values compatible with the ACT observations or cannot sustain the required number of e-folds under our initial-condition prescription. We observe that our potential can satisfy the ACT constraints for $\theta>0$ and $0.16\lesssim B\lesssim0.5$. The limit $B=0$ corresponds to natural inflation, which is already strongly disfavored by the Planck results. 

Interestingly, $f_{\rm min}$ (and thus ACT-compatible solutions) disappear at approximately $B>0.5$ for $A=1/2$, suggesting a characteristic boundary near $B=A$. This can be understood from Eqs.~(\ref{eq:potential-derivatives}) and (\ref{eq:stationary-degeneracy}) by noting that the relative amplitude of the slope contributed by the second sinusoidal term is proportional to $B/A$, such that for $B>A$ the higher-frequency term dominates the slope of the potential. Although local cancellations between the two contributions may still occur, they become increasingly confined to narrow field regions, making it difficult to form a sufficiently broad flat region capable of sustaining slow-roll inflation while keeping $r$ suppressed. This explains the disappearance of ACT-compatible solutions for $B\gtrsim A$.

Then, we select representative points at $B=0.33$ to illustrate how $f_{\rm min}$ varies as the shape of the potential changes with $\theta$ from $0$ to $3.021$. The corresponding potentials are shown in Fig.~\ref{fmin_potential_combined_Aeq0pt5}(b), while their trajectories in the $(n_s,r)$ plane are presented alongside the ACT contours in Fig.~\ref{ACT_ns_r_marked_Aeq0pt5}.
Starting from point {\circled{1}} at $\theta=0$, the field value at horizon exit, between the
global minimum and the first local maximum, generally lies in
a region where the potential is too negatively curved.
This results in a value of $n_s$ that is lower (more red-tilted)
than preferred by ACT.
Increasing $f$ can raise $n_s$, but the field value at horizon
exit also moves into a steeper region of the potential.
Consequently, $r\propto (V_{\phi}/V)^2$ increases, moving the
prediction away from the ACT contour, as shown in
Fig.~\ref{ACT_ns_r_marked_Aeq0pt5}.

As $\theta$ increases from point $\circled{2}$ to $\circled{4}$, we see flat regions open up that allow successful inflation at small $f_{\rm min}$. However, these flat regions become progressively steeper with increasing $\theta$, such that a higher $f_{\rm min}$ becomes necessary for the predicted $(n_s,r)$ to remain within the observational constraints. These viable points also occur at progressively higher values of $r$. A local minimum eventually develops at around $\theta = 3.021$, accompanied by the emergence of a new neighboring local maximum. This newly emerged maximum lies much closer to the global minimum in dimensionless field space than the previously relevant local maximum. The resulting dimensionless interval $\Delta X$ available before the end of inflation is much smaller. Since the physical field excursion is $\Delta\phi=f\Delta X$, the available physical field range also becomes too restricted to sustain the required number of e-folds of slow-roll inflation while remaining consistent with the ACT constraints, for the limited range of $f$ we are considering, i.e. $f\leq20\,M_{\rm Pl}$.

\subsection{Period ratio $A=1/4$}
\label{sec:scan-quarter}

\begin{figure*}[ht]
    \centering
    \includegraphics[width=\linewidth]{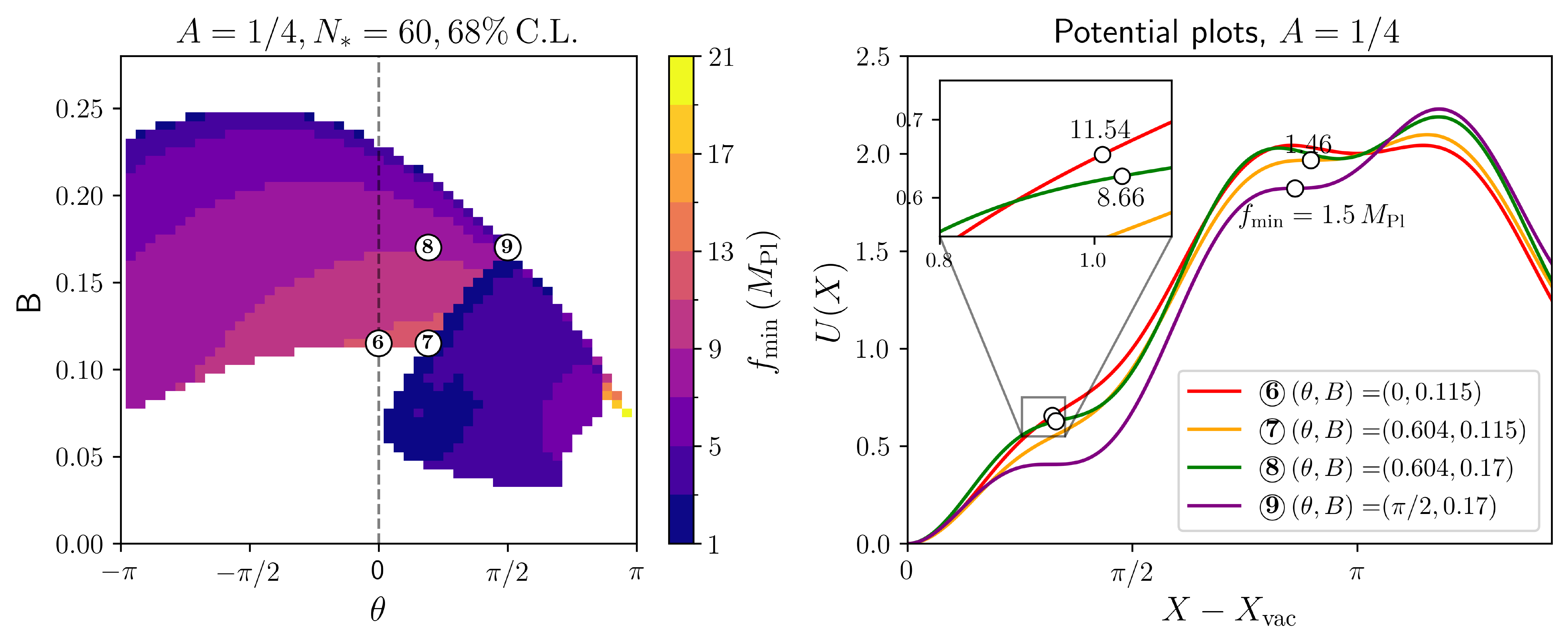}
    \caption{
        (a) Heatmap of minimum $f$ for different $(\theta,B)$ pairs at $A=1/4$, for which the predicted $(n_s,r)$ at $N_*=60$ first enter the ACT $68\%$ C.L. region. Representative points are selected and numbered. (b) Corresponding potentials for the numbered points in (a), with the field values at horizon crossing for $N_*=60$ and $f=f_{\rm min}$ indicated.  
    }
 \label{fmin_potential_combined_Aeq0pt25}
\end{figure*}

\begin{figure}[ht!]
    \centering
    \includegraphics[width=\linewidth]{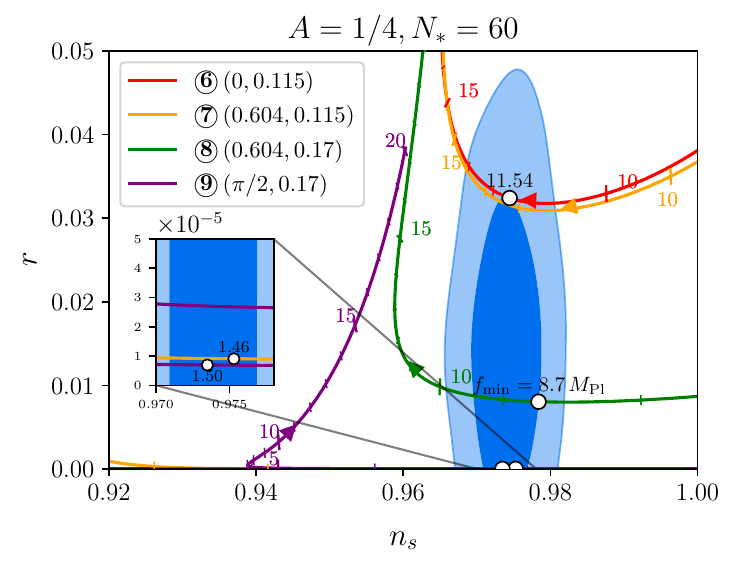}
    \caption{
    Same as Fig.~\ref{ACT_ns_r_marked_Aeq0pt5}, but for $A=1/4$. The legend shows the $(\theta,B)$ corresponding to the numbered parameter points in Fig.~\ref{fmin_potential_combined_Aeq0pt25}(a). The $f$-trajectories of \circled{7} and \circled{9} are shown in greater detail in the inset.
    }
    \label{ACT_ns_r_Aeq0pt25_withfscale}
\end{figure}

Similarly, we obtained Fig.~\ref{fmin_potential_combined_Aeq0pt25}(a) for $A=1/4$. Again, there exists a characteristic boundary at $B\simeq A=0.25$ as a consequence of Eqs.~(\ref{eq:potential-derivatives}) and (\ref{eq:stationary-degeneracy}). Compared with the $A=1/2$ case, the ACT-compatible region for $A=1/4$ extends over a wider range of $\theta$, including the negative values. This may be understood from the larger number of oscillations in the second cosine term within one period of the first. The resulting potential thus contains more locations where the contributions to the slope and curvature partially cancel, providing additional candidate flat regions that can support ACT-compatible slow-roll inflation. 

On the other hand, we note that the apparent asymmetry of $f_{\rm min}$ under $\theta\xrightarrow[]{}-\theta$
visible in both Fig.~\ref{fmin_potential_combined_Aeq0pt5}(a) and Fig.~\ref{fmin_potential_combined_Aeq0pt25}(a) arises from our choice of inflationary trajectory, as outlined at the beginning of this section. For a fixed $\theta$, trajectories approaching the minimum from opposite directions generally have different horizon-exit field values and therefore different inflationary observables. Owing to the symmetry $V(\phi;\theta)=V(-\phi;-\theta)$, the trajectory on the opposite side at $\theta$ is thus represented by the reflected trajectory at $-\theta$, allowing us to cover all distinct inflationary trajectories allowed by each potential.

Notably, $f_{\rm min}$ decreases sharply as $\theta$ crosses a particular value, which is especially apparent in the region $0\lesssim\theta\lesssim \pi/2$. This behavior can be understood from the evolution of the potential structure. At the representative point $\circled{6}$ where $f_{\rm min}=11.54\,M_{\rm Pl}$, the pivot scale exits the horizon on the relatively flat region near the vacuum, while the feature around $X-X_{\rm vac}=\pi$ corresponds to a local minimum. As $\theta$ is increased, this local minimum gradually develops into a near-inflection point region. Once sufficiently flat, this region can itself support slow-roll inflation, allowing an alternative ACT-compatible trajectory at a substantially smaller value of $f$. This is illustrated by the potential at $\circled{7}$ in Fig.~\ref{fmin_potential_combined_Aeq0pt25}(b), for which the horizon exit-point lies near the newly formed flat region and $f_{\rm min}$ decreases to $1.46\,M_{\rm Pl}$.

At fixed $\theta$, increasing $B$ causes a local minimum to re-emerge around this near-inflection-point structure. The ACT-compatible trajectory is then again confined to the flatter region closer to the vacuum, resulting in an increase in $f_{\rm min}$. A similar mechanism occurs near the representative point $\circled{9}$ at $\theta=\pi/2$, where the appearance of another near-inflection point region leads to a further decrease in $f_{\rm min}$.

\subsection{Hilltop and near-inflection-point regimes}
\label{subsec: hilltop inf}
We now turn to two representative phase regimes that illustrate how the interference between the two cosine terms produces ACT-compatible inflation: a slightly phase shifted quartic hilltop with $0<\theta\ll1$, and the configuration at $\theta=\pi$. The former demonstrates how ACT-compatible slow-roll inflation can be realized at sub-Planckian $f$, while the latter gives rise to different potential structures for $A=1/2$ and $A=1/4$, including a near-inflection-point slow-roll regime.

\subsubsection{Small relative phase shift $\theta$} \label{sec:scan-small-shift}
At $\theta=0$ and $B=A^2$, the expansion of the potential around the local maximum begins at quartic order, realizing quartic hilltop inflation \cite{Czerny:2014wza,Daido:2017wwb}. In the small-field limit with $f \ll M_{\rm Pl}$,
the scalar spectral index is approximately given by
\be
n_s\simeq 1-\frac{3}{N_*},
\ee
yielding $n_s\simeq 0.95$ and $0.94$ for $N_*=60$ and $50$,
respectively. Both values are significantly lower than that preferred by ACT and are therefore strongly disfavored.\footnote{The agreement with observations does not significantly improve for larger $f$.}
However, previous studies have shown that a small nonzero phase shift, $0<\theta\ll1$, introduces a linear term in the expansion around the hilltop~\cite{Takahashi:2013cxa,Daido:2017wwb,Takahashi:2019qmh}. The resulting nonzero first derivative term thus tilts the otherwise flat plateau and shifts the position of horizon exit closer to the potential maximum, thereby increasing the predicted value of $n_s$. Motivated by this idea, we examine the region around $B=A^2$ for small phase shifts in the range of $\theta\in[-\pi/32,\pi/32]$.

\begin{figure*}[ht]
    \centering
    \includegraphics[width=\linewidth]{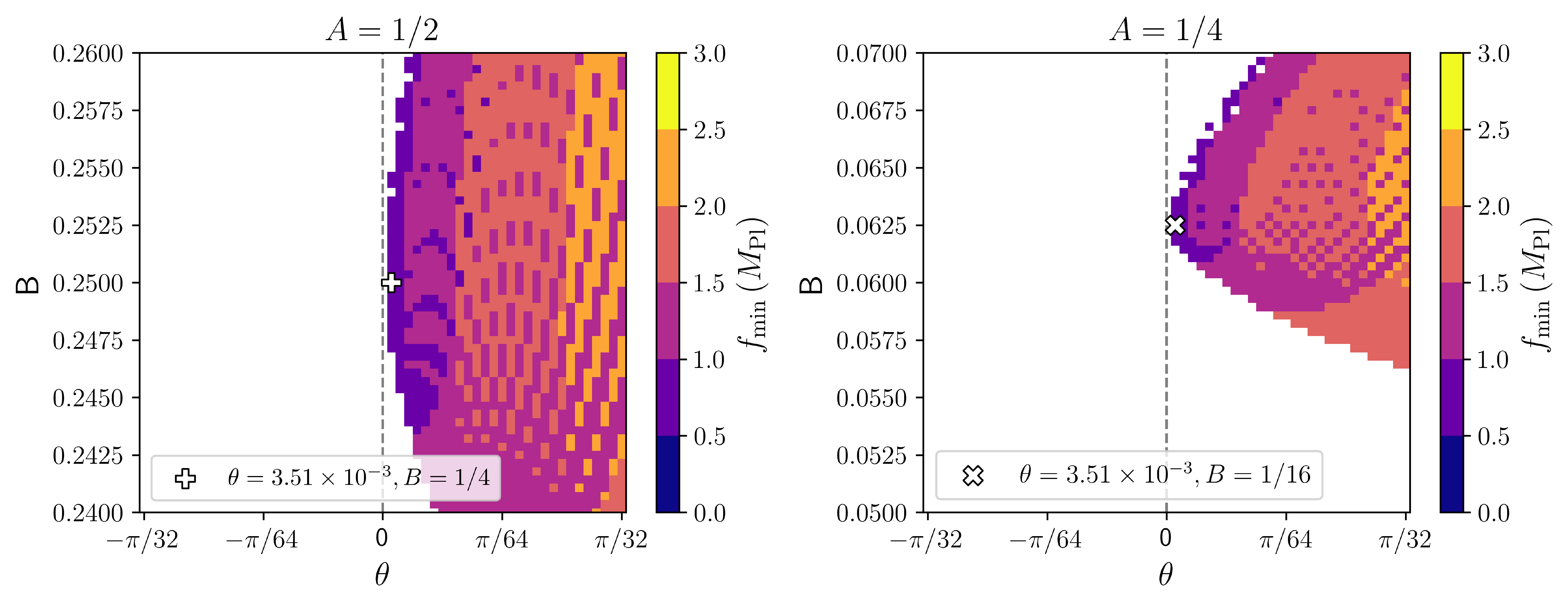}
    \caption{
        Heatmap of minimum $f$ for different $(\theta,B)$-pairs for small phase shift $\theta$ at (a) $A=1/2$ and (b) $A=1/4$, for which the predicted $(n_s,r)$ at $N_*=60$ first enter the ACT $68\%$ C.L. region. Points at $B=A^2$ are selected with the corresponding potentials shown in Fig.~\ref{potential_plot_both_subplanckian_withInset}.
    }
    \label{fmin_small_theta}
\end{figure*}

\begin{figure}[ht]
    \centering
    \includegraphics[width=
    \linewidth]{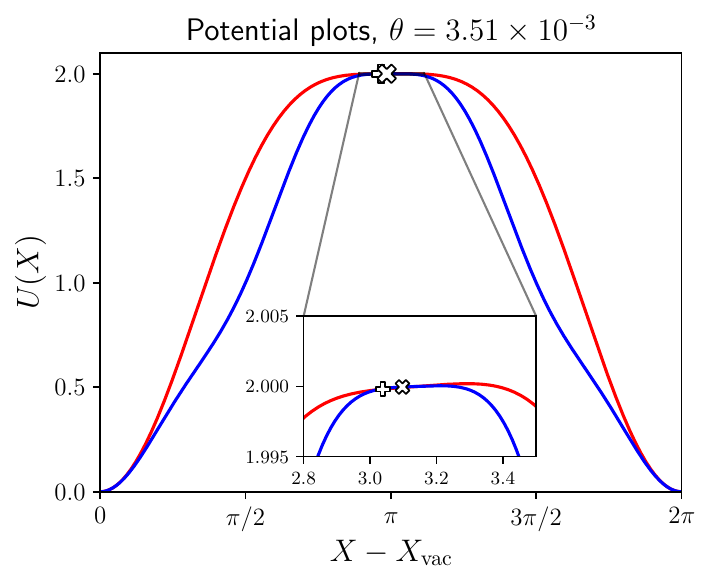}
    \caption{
    Dimensionless inflaton potentials $U(X)$ for $(A=1/2,B=1/4)$ (red solid line) and $(A=1/4,B=1/16)$ (blue solid line), corresponding to the parameters of the selected points in Fig.~\ref{fmin_small_theta}. Markers indicate the dimensionless field values (shifted) at which the pivot scale exits the horizon $N_{\ast}=60$ when $f=f_{\rm min}=0.75\,M_{\rm Pl}$ in both cases. The inset magnifies the region near the flat plateau, illustrating a modified (slightly-tilted) hilltop inflation.
    } 
    \label{potential_plot_both_subplanckian_withInset}
\end{figure}

Fig.~\ref{fmin_small_theta} shows $f_{\rm min}$, the smallest value of $f$ falling within the ACT 68\% C.L. region for $A=1/2$ and $A=1/4$, in the vicinity of $B=A^2$ for small phase shifts $\theta$. We find that ACT-compatible sub-Planckian values of $f_{\rm min}$ become possible due to the formation of slightly tilted flat regions with small slope and curvature, arising from the local partial cancellation between the two cosine terms in the potential.\footnote{
Such modified quartic hilltop inflation can be consistent with the ACT even for $f \ll M_{\rm Pl}$~\cite{Daido:2017wwb,Takahashi:2019qmh}.
} The corresponding potentials shown in Fig.~\ref{potential_plot_both_subplanckian_withInset} can therefore sustain a sufficient number of e-folds while producing $(n_s,r)$ values compatible with the ACT constraints. 

\subsubsection{Reflection-symmetric case: $\theta=\pi$}
\label{sec:scan-reflection}

We next consider $\theta=\pi$, for which the potential is
reflection symmetric. This symmetry can relate two distinct
degenerate minima, while the shape of the barrier between them
determines whether nearly flat shoulders develop.
We examine the coexistence of these features as a possible
realization of topological inflation.

For $\theta=\pi$, the potential becomes
$U(X)=c-\cos X+B\cos(X/A)$.
Changes in its stationary-point structure occur when
$U_X=U_{XX}=0$. Eliminating $B$ gives
\begin{equation}
\sin X_s\cos(X_s/A)
-A\cos X_s\sin(X_s/A)=0.
\label{eq:derivatives_theta_pi}
\end{equation}
For the solutions relevant to $A=1/2$ and $A=1/4$, the
corresponding amplitude ratio is
\begin{equation}
B=A^2\frac{\cos X_s}{\cos(X_s/A)}.
\label{eq:B_constraint_theta_pi}
\end{equation}
For $A=1/2$, Eq.~\eqref{eq:derivatives_theta_pi} reduces to
$\sin^3X_s=0$.
Within $0\leq X_s\leq\pi$, the only solution with $B>0$
is $X_s=0$ at $B=1/4$.
Since $U_{XXX}(0)=0$ and $U_{XXXX}(0)=3$ at this point,
the origin is a quartic minimum.
For $B>1/4$, it becomes a local maximum, and two degenerate
minima appear at
$X=\pm\arccos[1/(4B)]$ modulo $2\pi$.
The maximum at $X=\pi$ persists throughout this transition.

For $A=1/4$, Eq.~\eqref{eq:derivatives_theta_pi} instead reduces to
\begin{equation}
\sin^3X_s\left(1-6\cos^2X_s\right)=0.
\label{eq:critical-factorization-quarter}
\end{equation}
The solution at the origin now occurs at $B=1/16$, where
$U_{XXXX}(0)=15$.
Increasing $B$ through this value again splits the quartic
minimum into two reflection-related degenerate minima.
In addition, the nonzero solutions with $B>0$ give
\begin{align}
B_{\rm crit}&=\frac{3\sqrt{6}}{32}
\simeq0.22964,
\notag\\
X_s&=\arccos\left(-\frac{1}{\sqrt{6}}\right)
\simeq1.99133.
\label{eq:exact-shoulder-critical}
\end{align}
At this critical point,
$U_{XXX}(X_s)=5\sqrt{30}/2\neq0$, so $X_s$ is a stationary
inflection point.
Reflection symmetry gives a second stationary inflection point
at $2\pi-X_s\simeq4.29185$.

\begin{figure*}[ht]
    \centering
    \includegraphics[width=\linewidth]{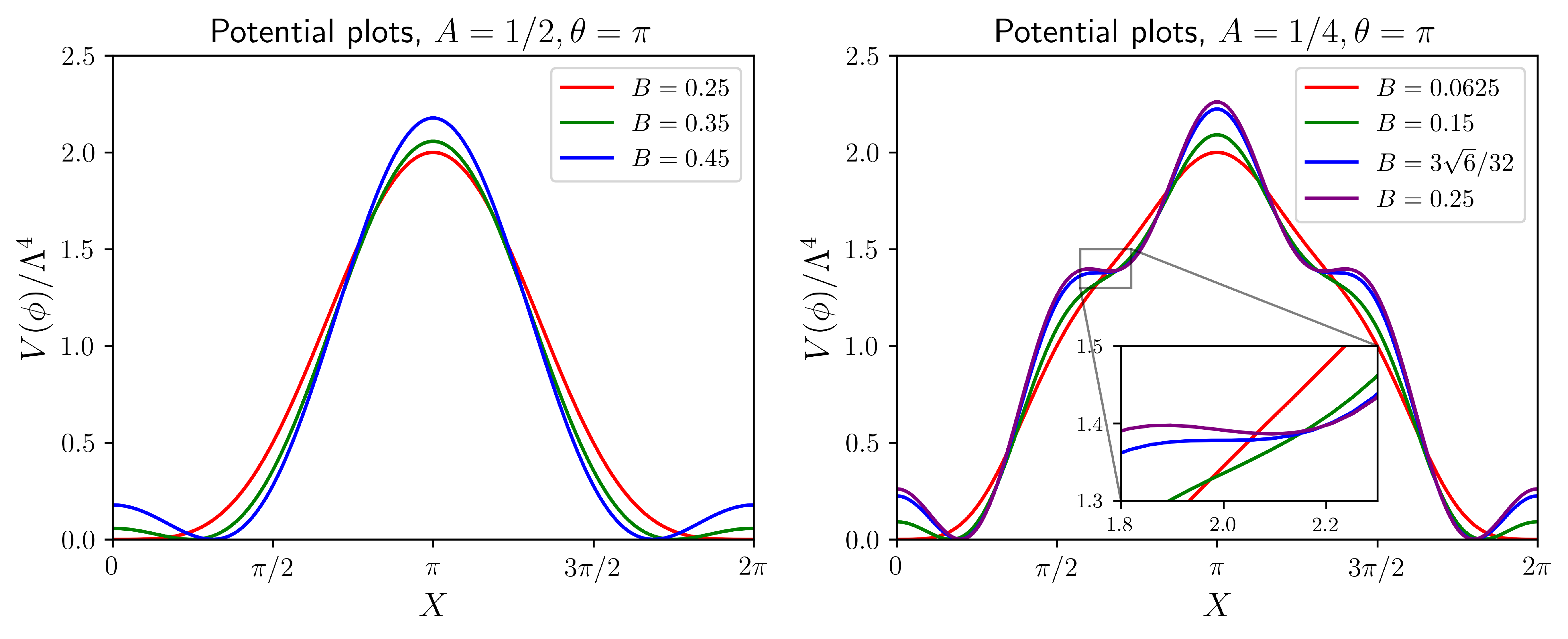}
    \caption{
        Dimensionless potentials $U(X)=V(\phi)/\Lambda^4$
        at $\theta=\pi$ for (a) $A=1/2$ and (b) $A=1/4$,
        plotted against $X=\phi/f$, instead of $X-X_{\rm vac}$.
        The quartic minimum at the origin splits into two
        degenerate minima when $B$ exceeds $A^2$.
        For $A=1/4$, a pair of stationary inflection points
        additionally appears at
        $B_{\rm crit}=3\sqrt{6}/32\simeq0.22964$.
        The inset shows the shoulder becoming flat as this
        critical value is approached, and the additional
        local extrema for $B>B_{\rm crit}$.
    }
    \label{potential_plot_both_thetaPi}
\end{figure*}

As $B$ approaches $B_{\rm crit}$ from below, nearly flat
shoulders develop around these two field values while retaining
a nonzero slope.
They can therefore support slow-roll evolution toward the
global minima, as discussed in Sec.~\ref{sec:inflection-regime}.
For $B>B_{\rm crit}$, a local minimum--maximum pair appears
near each shoulder, and the new local minimum can trap a field
rolling from the central maximum.
Fig.~\ref{potential_plot_both_thetaPi} illustrates the
different stationary-point structures for the two period ratios.

For $A=1/4$ and $1/16<B<B_{\rm crit}$ sufficiently close to
$B_{\rm crit}$, the potential thus combines two degenerate
minima with nearly flat shoulders on the barrier between them.
We focus on walls interpolating between these minima through
$X=\pi$.
Their cores are centered on this maximum, while CMB
perturbations can be generated along either shoulder.
In the next section, we examine wall formation and the
conditions for topological inflation in this regime and in
the small-$\theta$ regime.


\section{Topological inflation}
\label{sec:topological}
We now examine whether the CMB-compatible potentials identified
in Sec.~\ref{sec:scan} also admit a topological origin of
inflation. We first describe a symmetry-breaking history that
produces axion domain walls, then introduce diagnostics for
inflation in their cores, and finally apply these diagnostics
to the small-$\theta$ and reflection-symmetric regimes.

\subsection{Axion vacuum structure and domain-wall formation}
\label{sec:vacua}

An attractive feature of topological inflation is that the
initial field configuration required for inflation can arise
naturally from a cosmological symmetry-breaking phase
transition. In particular, axion domain walls provide a
natural setting for this mechanism, since different
degenerate vacua can be populated in causally disconnected
regions of the Universe.

To illustrate this possibility, consider a complex scalar
field $\Phi=|\Phi|{\rm e}^{i\phi/F}$ with an approximate
global $U(1)$ symmetry. We assume that the Universe initially
resides in the symmetric phase, with $\langle\Phi\rangle=0$,
as may occur due to thermal effects in the early Universe.
As the Universe evolves, the $U(1)$ symmetry is spontaneously
broken and $\Phi$ acquires a nonzero radial vacuum expectation
value. The axion field then emerges as the angular
component, which takes random initial values in
causally disconnected regions.

As the field interpolates continuously between these regions,
configurations with nonzero winding around the $U(1)$ vacuum
manifold give rise to cosmic strings.
At a later stage, explicit symmetry-breaking effects become
dynamically relevant and generate a periodic axion potential
with multiple degenerate minima. Different regions can then
evolve toward different vacua, resulting in the formation of
axion domain walls attached to the pre-existing strings.
If the wall cores are sufficiently broad and dominated by
potential energy, they can undergo topological inflation
without requiring finely tuned initial field values.

\subsection{Conditions and diagnostics for topological inflation}
\label{sec:wall-conditions}
We consider two main requirements for identifying configurations for topological inflation. First, distinct degenerate vacua have to be populated in spatially-separated regions, forcing the field to interpolate them. The symmetry-breaking history described above provides a natural mechanism for realizing such configurations.

Secondly, the domain wall must be sufficiently thick to undergo inflation. A heuristic estimate for the required thickness was proposed by Vilenkin \cite{Vilenkin:1994pv}:
\be \label{eq:vilenkin-thickness}
\delta_{\rm wall} \gtrsim H_{\rm inf}^{-1},
\ee
where $H_{\rm inf}$ is the Hubble parameter during inflation, $H_{\rm inf}^2 \simeq V(\phi_{\rm top})/3M_{\rm Pl}^2$. Here $\phi_{\rm top}$ denotes the axion field value at the potential maximum.

We estimate the wall thickness from the corresponding
flat-spacetime static profile, for which
$\frac12(d\phi/dz)^2=V(\phi)$.
We define the width using the central $80\%$ of the
field interval between the two degenerate vacua,
\begin{equation}
\delta_{\rm wall}
 = \int_{\phi_{10}}^{\phi_{90}}
   \frac{d\phi}{\sqrt{2V(\phi)}} .
\end{equation}
We use Eq.~(\ref{eq:vilenkin-thickness}) to identify
candidate configurations for topological inflation.

As a complementary diagnostic, we also consider the existence of
stationary, self-gravitating domain-wall solutions, following
Bonjour et al.~\cite{Bonjour:1999kz}.
They considered plane-symmetric walls whose scalar field
profile across the wall is time-independent, while the
geometry along the wall may expand.
For a reflection-symmetric potential about $X_{\rm top} = \phi_{\rm top}/f$,
with $|U_{XX}|$ maximized at $X_{\rm top}$, they showed that
no regular stationary wall solution exists if~\footnote{
In the conventions of Ref.~\cite{Bonjour:1999kz}, the field
scale is $\eta=f/\sqrt{2}$, giving
$\epsilon_g=8\pi G\eta^2=f^2/(2M_{\rm Pl}^2)$.
}
\begin{equation}
\frac{M_{\rm Pl}^2}{f^2}
\frac{|U_{XX}(X_{\rm top})|}{U(X_{\rm top})}<1.
\label{eq:no-static-wall-limit}
\end{equation}
Thus, a sufficiently small barrier curvature relative
to the Hubble scale precludes a stationary wall profile.
In the quartic hilltop limit, however,
$U_{XX}(X_{\rm top})=0$, while $|U_{XX}|$ becomes
nonzero away from the top.
The assumption that $|U_{XX}|$ is maximized at the
barrier top is therefore violated, and the criterion
does not apply to this limit.
By contrast, for the reflection-symmetric configurations
at $\theta=\pi$ considered below, $|U_{XX}|$ is maximized
at $X_{\rm top}=\pi$ for both $A=1/2$ and $A=1/4$,
so the criterion is applicable.

\subsection{Parameter regions compatible with the CMB}
\label{sec:topological-regions}

We now combine these wall diagnostics with the CMB constraints
on $(n_s,r)$. For the small-$\theta$ configurations,
we use the wall-width criterion.
For the reflection-symmetric configurations at $\theta=\pi$,
we additionally apply the bound in
Eq.~\eqref{eq:no-static-wall-limit}.

\subsubsection{Small relative phase shift $\theta$}
\begin{figure*}[ht]
    \centering
    \includegraphics[width=\linewidth]{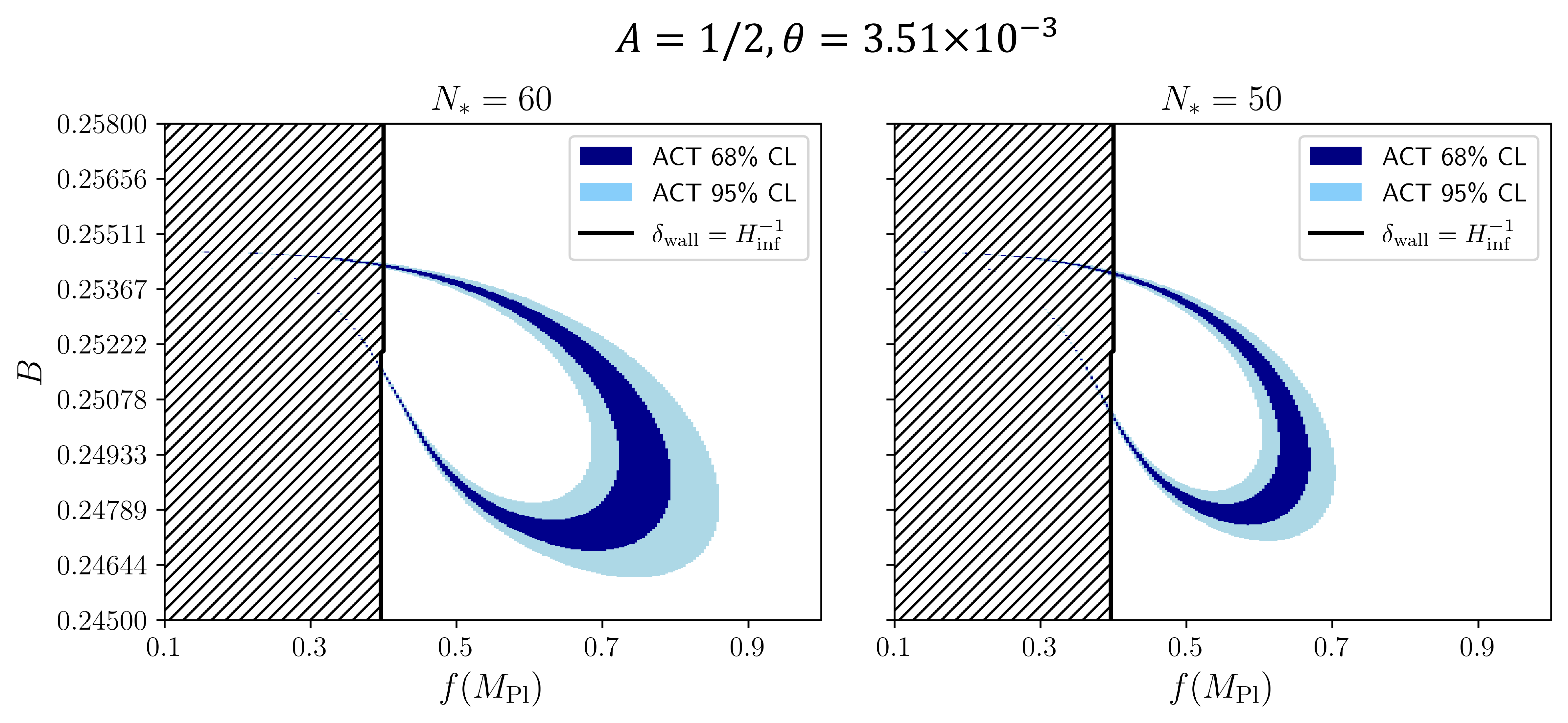}
    \caption{
CMB-compatible regions in the $(f,B)$ plane at
$A=1/2$ and $\theta=3.51\times10^{-3}$,
for $N_*=60$ (left) and $N_*=50$ (right).
The dark and light blue regions satisfy the 68\% and 95\%
C.L. constraints, respectively.
The solid black line denotes $H_{\rm inf}\delta_{\rm wall}=1$,
and the hatched region corresponds to
$H_{\rm inf}\delta_{\rm wall}<1$.
    }
    \label{plot_Aeq0pt5_thetaSmall_insideACT_thickness}
\end{figure*}

\begin{figure*}[ht]
    \centering
    \includegraphics[width=\linewidth]{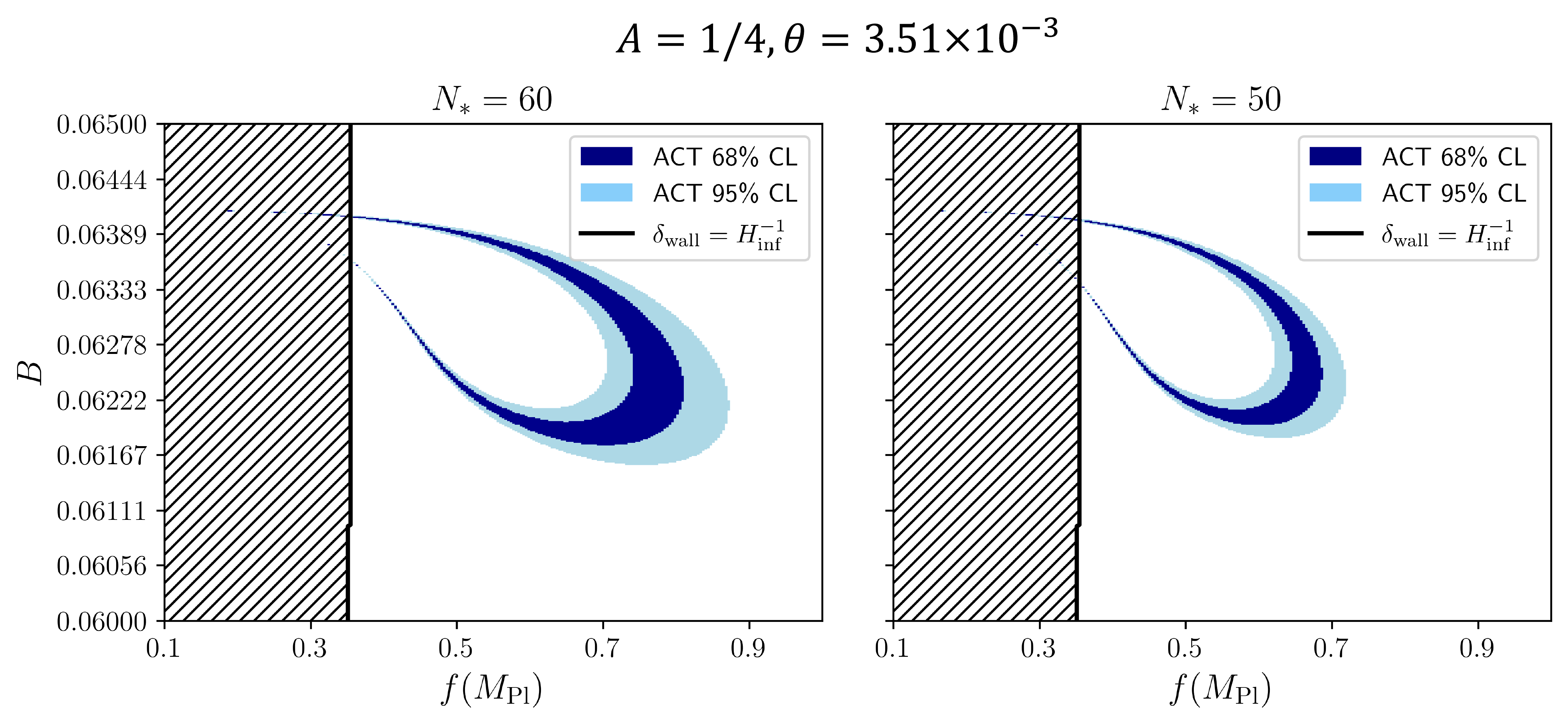}
    \caption{
        Same as Fig.~\ref{plot_Aeq0pt5_thetaSmall_insideACT_thickness} but with $A=1/4$.
    }
    \label{plot_Aeq0pt25_thetaSmall_insideACT_thickness}
\end{figure*}

For the small-$\theta$ configurations considered here, the
potential has a single global minimum within each period
$2\pi f$.
As shown in Sec.~\ref{sec:scan-small-shift},
small phase shifts near $B=A^2$ allow ACT-compatible
inflation with sub-Planckian $f$.
We therefore fix the representative phase shift
$\theta=3.51\times10^{-3}$ and scan the $(f,B)$ parameter
space around $B=A^2$ for both $A=1/2$ and $A=1/4$,
restricting to $f\leq M_{\rm Pl}$.
The results are shown in
Figs.~\ref{plot_Aeq0pt5_thetaSmall_insideACT_thickness}
and \ref{plot_Aeq0pt25_thetaSmall_insideACT_thickness},
respectively.
The dark and light blue regions correspond to predictions
within the ACT $68\%$ and $95\%$ confidence regions,
while the hatched regions fail the wall-width criterion
in Eq.~\eqref{eq:vilenkin-thickness}.
For both period ratios, we find CMB-compatible regions
that satisfy this criterion, providing candidate
configurations for topological inflation.

In both $A=1/2$ and $A=1/4$ cases, the ACT-compatible bands for $N_*=50$ are displaced towards smaller $f$ and become narrower. 
The displacement can be understood from the relation of $N_*$ under slow-roll approximation, given by
\be
N_*\simeq \frac{f^2}{M_{\rm Pl}^2}\bigg|\int^{X_*}_{X_{\rm end}}\frac{U}{U_X}\,dX\bigg|.
\ee
Writing the dimensionless integral as 
\be
\label{eq:fscale}
\mathcal{J}(B;X_*,X_{\rm end})\equiv\bigg|\int^{X_*}_{X_{\rm end}}\frac{U}{U_X}\,dx\bigg|,
\ee
we have
\be \label{f_N_relation}
f\simeq M_{\rm Pl}\sqrt{\frac{N_*}{\mathcal{J}}}.
\ee
Since the horizon-exit field values $X_*$ for $N_*=60$ and $N_*=50$ lie in approximately the same flat region, Eq.~(\ref{eq:fscale}) suggests the scaling
\be
f\propto\sqrt{N}.
\ee
Therefore, the overall displacement towards smaller $f$ is qualitatively consistent with the approximate scaling $f_{50}/f_{60}\sim\sqrt{50/60}\simeq0.913$. Meanwhile, the reduced width of the band for $N_*=50$ reflects the shift of the horizon exit nearer to the end of inflation, such that only a more restricted range of $(f,B)$ produces $(n_s,r)$ within the ACT region.

\subsubsection{Reflection-symmetric case: $\theta=\pi$}
\label{sec:reflection-symmetric-potential}

We next consider the reflection-symmetric potentials studied
in Sec.~\ref{sec:scan-reflection}. For both $A=1/2$ and
$A=1/4$, the origin is a quadratic minimum for $B<A^2$,
a quartic minimum at $B=A^2$, and a local maximum for
$B>A^2$. In the latter case, two reflection-related,
degenerate global minima appear within each potential
period.

We focus on the wall branch passing through $X=\pi$.
Writing the positive global minimum as $X_v$, the endpoints
of this branch are
\begin{equation}
X_-=X_v,\qquad X_+=2\pi-X_v,
\label{eq:reflection-wall-path}
\end{equation}
with $X_v=0$ for $B\leq A^2$.
For $B>A^2$, the two endpoints correspond to distinct
degenerate vacua related by reflection symmetry.

For the chosen period ratios, the potential
$U(X)=c-\cos X+B\cos(X/A)$ is reflection symmetric about
$X=\pi$. Moreover,
\begin{align}
|U_{XX}(X)|
&=\left|\cos X-\frac{B}{A^2}\cos(X/A)\right|
\notag\\
&\leq 1+\frac{B}{A^2}
=|U_{XX}(\pi)|.
\label{eq:reflection-curvature-bound}
\end{align}
The assumptions underlying Eq.~\eqref{eq:no-static-wall-limit}
are therefore satisfied. Using $U(\pi)=c+1+B$, the sufficient
no-static-wall bound becomes
\begin{equation}
\frac{f}{M_{\rm Pl}}>
\sqrt{\frac{1+B/A^2}{c+1+B}},
\label{eq:reflection-no-static-bound}
\end{equation}
where $c$ is fixed by $U(X_v)=0$.

For $A=1/2$, we scan
$B\in[0,0.5]$ and $f\in[0.1,10]M_{\rm Pl}$.
Within this range, we find no region for the wall branch
centered at $X=\pi$ that satisfies the adopted wall
diagnostics while producing $(n_s,r)$ within the ACT
confidence regions, for either $N_*=60$ or $50$.
This result is consistent with the absence of the nearly
flat shoulders that arise for $A=1/4$, as discussed in
Sec.~\ref{sec:scan-reflection}.

\begin{figure*}[ht]
    \centering
    \includegraphics[width=\linewidth]{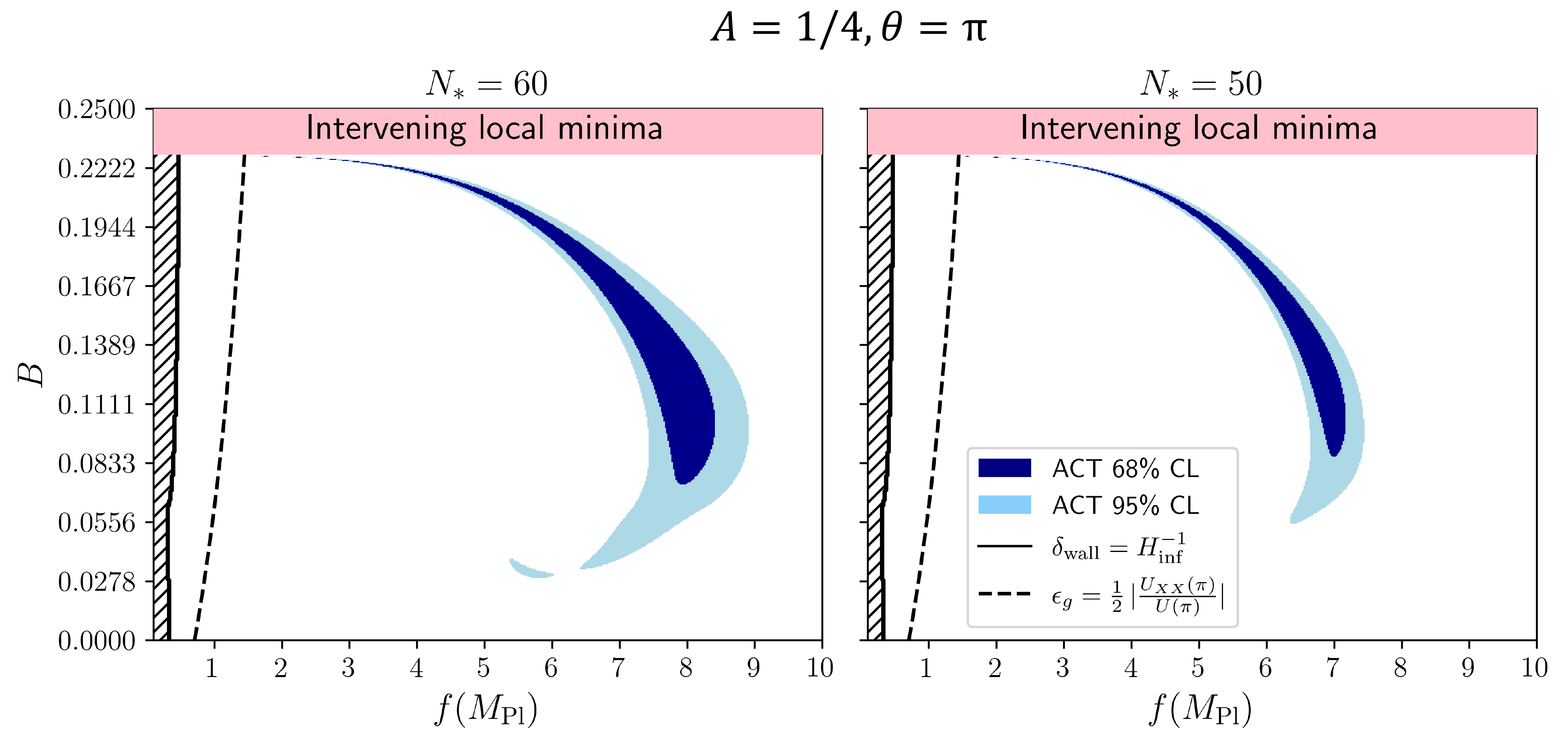}
    \caption{
        CMB-compatible regions in the $(f,B)$ plane at
        $A=1/4$ and $\theta=\pi$,
        for $N_*=60$ (left) and $N_*=50$ (right).The dashed line denotes the threshold $f$ value, above which the inequality in Eq.~(\ref{eq:no-static-wall-limit}) with $X_{\rm top}=\pi$ is satisfied, indicating that no static domain wall solution exists.  Above $B_{\rm crit}\simeq0.22964$, intervening local minima appear which can trap the inflaton.
    }
\label{plot_Aeq0pt25_thetaPi_insideACT_thickness}
\end{figure*}

For $A=1/4$, we scan
$B\in[0,0.25]$ and $f\in[0.1,10]M_{\rm Pl}$.
The results are shown in
Fig.~\ref{plot_Aeq0pt25_thetaPi_insideACT_thickness}.
In contrast to the $A=1/2$ case, CMB-compatible regions
satisfying both wall diagnostics occur for
$f\gtrsim2\,M_{\rm Pl}$.
The solid black line denotes $H_{\rm inf}\delta_{\rm wall}=1$,
while the dashed black line corresponds to equality in
Eq.~\eqref{eq:reflection-no-static-bound}.
The sufficient no-static-wall bound is satisfied on the
larger-$f$ side of the dashed line.

The $N_*=50$ band is again narrower and shifted toward
smaller $f$ than the $N_*=60$ band. For either value of
$N_*$, the bands bend toward smaller $f$ as $B$ approaches
$B_{\rm crit}$ from below. As shown in
Sec.~\ref{sec:scan-reflection}, the shoulders become
increasingly flat in this limit. The suppression of $|U_X|$
enhances the contribution of the shoulder to $\mathcal J$,
allowing the required number of e-folds to be obtained at
smaller $f$, as suggested by Eq.~\eqref{f_N_relation}.
For $B>B_{\rm crit}=3\sqrt{6}/32$, additional local minima
appear along the trajectories from $X=\pi$ and can trap
the inflaton. The candidate trajectories discussed here
therefore lie below this critical value.

In particular, the region with $A^2<B<B_{\rm crit}$
contains two physically distinct degenerate minima and,
near $B_{\rm crit}$, nearly flat shoulders between them.
The wall crosses the maximum at $X=\pi$, while the observable
slow-roll stage takes place along a shoulder. Thus the
potential structure supporting the wall and the local
slope and curvature controlling the CMB observables need
not coincide. The overlap of the CMB-compatible bands
with both wall diagnostics identifies reflection-symmetric
candidate configurations for topological inflation in the
near-inflection-point regime.

\section{Discussion and conclusions}
\label{sec:conclusions}

We have studied topological inflation in minimal multi-natural
inflation, with an axion potential consisting of two cosine terms.
For the representative period ratios $A=1/2$ and $A=1/4$, we
scanned the relative amplitude, phase, and axion decay constant
and identified regions consistent with the CMB constraints on
$n_s$ and $r$, including those from ACT DR6.
The allowed regions include both deformed hilltop potentials
and potentials with nearly flat shoulders approaching stationary
inflection points.

For small nonzero $\theta$ and $B$ close to $A^2$, the relative
phase introduces a linear term near the quartic hilltop.
This raises $n_s$ above the small-field quartic-hilltop prediction
$n_s \simeq 1-3/N_*$ and brings the predicted spectrum into
agreement with the observations.
For both period ratios, we find regions with $f<M_{\rm Pl}$
that satisfy the CMB constraints and the wall-width criterion
$\delta_{\rm wall}\gtrsim H_{\rm inf}^{-1}$
of Ref.~\cite{Vilenkin:1994pv}, with the wall thickness estimated
from the corresponding flat-spacetime static profile.
These regions provide candidate configurations for topological
inflation with a sub-Planckian axion decay constant.

A distinct regime appears at $A=1/4$ and $\theta=\pi$.
For $B$ just below $B_{\rm crit}=3\sqrt{6}/32$, the potential
has two reflection-related degenerate minima and nearly flat
shoulders on either side of the central maximum at $X=\pi$.
We find CMB-compatible regions that satisfy both the wall-width
criterion and the sufficient bound for the absence of stationary,
self-gravitating wall solutions derived by Bonjour
et al.~\cite{Bonjour:1999kz}.
This bound applies because the potential is reflection symmetric
about $X=\pi$ and $|U_{XX}|$ is maximal there.
The wall core is centered on the potential maximum,
while observable perturbations are generated during
slow roll along a nearly flat shoulder.
This separation allows the CMB constraints and both
wall diagnostics to be satisfied simultaneously.

The axion interpretation also provides a natural
mechanism for forming the required domain walls.
Spontaneous breaking of the underlying U(1) symmetry
produces cosmic strings, while subsequent explicit
symmetry breaking gives rise to domain walls attached
to them.
The radial potential may also drive an earlier stage
of topological inflation in string cores~\cite{Vilenkin:1994pv},
preceding inflation associated with the angular field.

Our assessment of topological inflation is based on simple
criteria for the wall thickness and the existence of stationary
wall solutions.
A next step is to solve the time-dependent, coupled scalar-field
and Einstein equations for these potentials.
Such a calculation would determine the dynamical boundary
between inflating and noninflating walls as a function of the
potential parameters and initial data.
It would also test the wall-width estimate for asymmetric
and nonplanar configurations beyond the assumptions of the
stationary-wall bound.
For the near-inflection-point branch, the same evolution would
clarify how regions leaving the wall core enter the slow-roll
trajectory along a shoulder.
We leave this numerical study for future work.

The observational tension encountered in the minimal quartic
double-well model therefore need not extend to topological
inflation in general.
Two harmonics already suffice to satisfy the CMB constraints
together with the wall criteria.
In particular, the near-inflection-point branch offers a route
to generating the observed perturbations during a slow-roll
phase whose initial conditions have a topological origin.

\acknowledgments
This work was supported by JSPS KAKENHI Grant Numbers
25H02165 and 26K00695 (F.T.), by the World Premier International
Research Center Initiative (WPI), MEXT, Japan, and by COST Action
COSMIC WISPers CA21106.


\bibliography{references}
\end{document}